\documentclass[aps,prb,preprint,groupedaddress,citeautoscript]{revtex4}

\usepackage{amssymb}
\usepackage{graphicx}
\usepackage{epstopdf}

\begin{document}

\preprint{J. Chem. Phys. (in press)}

\title{Mesoscopic order and the dimensionality of long-range resonance energy transfer in supramolecular semiconductors}

\author{Cl\'{e}ment Daniel}
\author{Fran\c{c}ois Makereel}
\affiliation{Cavendish Laboratory, University of Cambridge, J.J.\ Thomson Avenue, Cambridge CH3~0HE, United Kingdom}

\author{Laura M.\ Herz}
\affiliation{Clarendon Laboratory, University of Oxford, Parks Road, Oxford OX1 3PU, United Kingdom}

\author{Freek J.\ M.\ Hoeben}
\author{Pascal Jonkheijm}
\author{Albertus P.\ H.\ J.\ Schenning}
\author{E.\ W.\ Meijer}
\affiliation{Laboratory of Macromolecular and Organic Chemistry, Eindhoven University of Technology, P.O. Box 513, 5600 MB Eindhoven, The Netherlands}

\author{Carlos Silva}
\email[Corresponding author. Electronic mail: ]{carlos.silva@umontreal.ca}
\affiliation{D\'{e}partement de Physique et Regroupement qu\'eb\'ecois sur les mat\'eriaux de pointe, Universit\'{e} de Montr\'{e}al, C.P.\ 6128, succ.\ centre-ville, Montr\'{e}al (Qu\'{e}bec) H3C 3J7, Canada}

\date{\today}

\begin{abstract}
We present time-resolved photoluminescence measurements on two series of oligo-\emph{p}-phenylenevinylene materials that self-assemble into supramolecular nanostructures with thermotropic reversibility in dodecane. One set of derivatives form chiral, helical stacks while the second set form less organised, `frustrated' stacks. Here we study the effects of supramolecular organisation on the resonance energy transfer rates. We measure these rates in nanoassemblies formed with mixed blends of oligomers and compare them with the rates predicted by F\"{o}rster theory. Our results and analysis show that control of supramolecular order in the nanometre lengthscale has a dominant effect on the efficiency and dimensionality of resonance energy transfer. 
\end{abstract}

\maketitle

\section{Introduction}

One of the most attractive properties of $\pi$-conjugated polymers as active materials in optoelectronic applications, from a processing point of view, is that they are soluble in common solvents and can therefore be cast using techniques such as ink-jet printing.\cite{Sirringhaus00,Kawase01} It is increasingly evident that controlling three-dimensional intermolecular structure in the solid state is essential to optimise the electronic properties of polymeric semiconductors. For example, the field-effect carrier mobility is orders of magnitude higher if conjugated polymer chains adopt lamellar intermolecular structure with chains aligned orthogonal to the transport direction,\cite{Sirringhaus99} characterized by weakly-coupled H-aggregates~\cite{Spano05,Clark07}. Supramolecular chemistry is a promising approach to achieve three-dimensional control of intermolecular interactions.\cite{Lehn95,Hoeben05b} This approach allows the design of extended complex structures built through the hierarchically ordered assembly of elementary building blocks in solution prior to the casting process using non-covalent interactions such as hydrogen bonding and $\pi$--$\pi$ interactions. Exploiting the technological interest in solution processing, supramolecular architectures may be assembled in solution and readily transferred to the solid state, providing the optoelectronic properties of polymeric semiconductors with a tailored three-dimensional structure to enhance a specific property such as charge mobilities.\cite{vandeCraats99,Duan02}

In bulk polymeric semiconductors, inter-chromophore coupling, where chromophores consist of $\pi$-conjugated segments within a chain, can have profound effects on the optoelectronics properties.\cite{Rothberg02,Schwartz03,Spano06} An important one is to facilitate both intrinsic~\cite{Spano05,Clark07,Cornil98} and extrinsic~\cite{Beljonne02,Hennebicq05,Herz04} luminescence quenching processes. Intrinsic quenching is related to dispersion of excitonic energy levels in an H-like aggregate and to modified internal conversion rates with respect to isolated chains. On the other hand, extrinsic processes can be enhanced by diffusion-limited quenching at either chemical or structural defects. These phenomena have significant effects on the photophysics even in the weak intermolecular coupling limit (when the intermolecular coupling is smaller than the intramolecular vibronic coupling). In this case, the exciton diffusion mechanism is incoherent hopping by resonance energy transfer (RET) between sites. In conjugated materials, either intermolecular (in polymer films) or intramolecular (in dilute polymer solution) RET is fundamental to describe exciton dynamics.\cite{Nguyen00a, Meskers00,Meskers01a,Beljonne02,Hennebicq05,Herz04} We are interested in developing an understanding of these phenomena in a model supramolecular system with controlled structural order compared to standard polymeric semiconductor systems (conjugated polymer films), and in doing so to contribute to the understanding of exciton dynamics in nanoscale systems.\cite{Scholes06} 

\begin{figure}
\includegraphics[width=12cm]{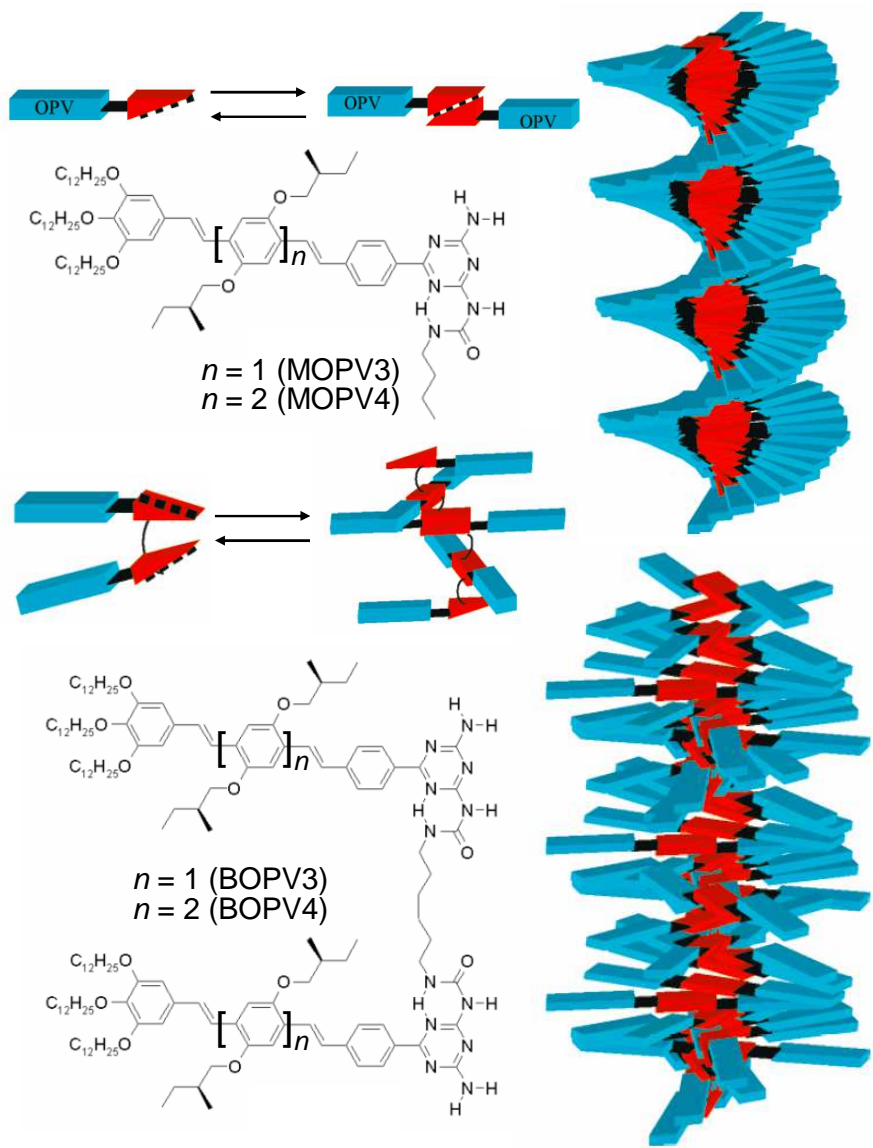}\\
\caption{Molecular structures of MOPV and BOPV derivatives and
schematic representation of the supramolecular structure in dodecane.}\label{FOMOPV}
\end{figure}

Here, we investigate RET kinetics in two pairs of oligo-\emph{p}-phenyl\-ene\-vinyl\-ene (OPV) derivatives  (see Fig.~\ref{FOMOPV}). MOPV and BOPV form dimers by hydrogen bonding in dodecane solution.\cite{Schenning01,Jonkheijm03,Jonkheijm05,Hoeben05} Solvophobic and $\pi$-$\pi$ interactions result in thermotropically reversible supramolecular assembly. These nanostructures have been characterised extensively by means of several techniques including circular dichroism measurements, neutron scattering, and scanning probe microscopies~\citep{Schenning01}. The cartoon shown in Fig.~\ref{FOMOPV} is actually an accurate picture of what these nanostructures are in solution. They may reach lengths of up to microns, whilst the diameter of the stacks corresponds to the length of the dimers. The intermolecular electronic coupling in the stacks is moderately strong compared to intramolecular vibronic coupling~\cite{Spano07}, resulting in red-shifted photoluminescence spectra (by up to 0.2\,eV) in MOPV stacks compared to MOPV solution. Similar shifts are observed in BOPV stacks, suggesting that the magnitude of intermolecular coupling is comparable in this system, although there is no supramolecular chirality. At the solution concentrations investigated here, MOPV undergoes a phase transition in the temperatures range between 50 and 70$^{\circ}$C . Due to its dimeric structure, BOPV forms a random coil supramolecular polymer in chloroform. In dodecane, the coils collapse to frustrated stacks, bringing the OPV units closer together. By raising the temperature, the distance between the OPVs increases but the result is the stretching of the frustrated stacks and not a complete break-up, as is the case in the MOPVs. This does not result in a well-defined phase transition in BOPVs, and we observe spectral changes from roughly 40$^{\circ}$C to 90$^{\circ}$C~\cite{Hoeben05}. In dodecane, MOPV assemblies are chiral with a small relative angle between oligomers and a small oligomer separation~\cite{Hoeben03}. On the other hand, the alkyl linking chains in the BOPV molecules hinder the packing and lead to more disordered, achiral frustrated-stack assemblies in dodecane~\cite{Schenning01,Jonkheijm03,Jonkheijm05}. This self-organisation allows us to study excitonic processes in various morphologies of isolated supramolecular nanostructures and to compare them with excited-state phenomena in dissolved oligomer solutions.

In previous work, we explored the extrinsic consequences of intermolecular coupling, namely diffusion-assisted exciton transfer and quenching and exciton bimolecular annihilation at high exciton densities~\cite{Herz03,Hoeben03,Daniel03,Daniel04,DanielDec04,Beljonne05,Chang06,Daniel07}.  We have demonstrated that RET between MOPV derivatives of different length (and exciton energies) is greatly enhanced by supramolecular assembly. At low MOPV4 mole fraction ($\lesssim 2$\%), isolated MOPV4 chromophores are incorporated into MOPV3 helical  assemblies as long as the solution is thermally cycled to dissolve and then re-assemble the stacks~\cite{Hoeben03}. Optical excitation of the blended structure results in efficient energy transfer from MOPV3 hosts to MOPV4 guests, with most of the transfer occurring over the first 100\,ps. Over this initial period the photoexcitation in the architecture, which is mostly composed of the donor oligomer, is highly mobile~\cite{Daniel07}. Energy transfer to the trap sites (the longer oligomer) is therefore mostly assisted by diffusion. The dominant interactions when such dynamics are important are close to nearest co-facial neighbour interactions. Once the excitation is no longer mobile in the donor phase, which occurs on timescales longer than 100\,ps, any residual energy transfer steps involve one-step transfer events in a \emph{static} donor-acceptor distribution. These would be over large average distances (in the order of nanometers at the acceptor concentrations considered in this discussion) and consequently over timescales that are long compared to the fast energy diffusion timescales.

Here we consider explicitly the long-time RET regime discussed above, involving an essentially static donor-acceptor distribution. The objective of this paper is to study the extrinsic consequences of chromophore packing and of morphology by measuring long-range RET between \emph{localised states} (i.e.\ when the excitation mobility is low) in MOPV and BOPV supramolecular architecture. These localised states have been found to comprise of two cofacial oligomers in MOPV nanostructures by circularly polarised absorption and emission studies and quantum chemical calculations~\cite{Spano07}, but are probably confined to a single oligomer in BOPV. We are particularly interested in exploring the correlation between supramolecular order and the dimensionality of RET. By this we mean that we are interested to probe whether or not inducing supramolecular order directs RET along a preferential axis in the types of chiral structures designed for this body of work. We find that in MOPV host nanostructures, one-dimensional RET dominates, but in more disordered BOPV nanostructures, the dimensionality of the RET process is higher. This is because the induced periodicity in the MOPV stack provides an essentially one-dimensional donor-acceptor distribution, while the distribution is les directed when structural disorder is more important. These results indicate the importance of the nanostructure morphology to the design of their electronic properties.

\section{Experimental Description}\label{ExpSec}
The synthesis of MOPV and BOPV derivatives has been described in detail elsewhere~\cite{Schenning01,Jonkheijm03,Hoeben05}. Materials were dissolved in anhydrous dodecane at concentrations
around $10^{-4}$\,M and then kept under inert atmosphere except during absorption measurements. For the blend measurements, the MOPV3 and BOPV3 concentrations were kept around $1.4 \times 10^{-4}$ and $0.8 \times 10^{-4}$\,M while the mole fractions of MOPV4 and BOPV4 were varied by titration from 0\% to 15\%. MOPV4 and BOPV4 were incorporated into MOPV3 and BOPV3 stacks by heating the solution to 80\,$^{\circ}$C after each titration to partly dissolve the stack, and then cooling the solution to a temperature well below the transition temperature for supramolecular assembly~\cite{Jonkheijm03,Hoeben05}, usually to 14\,$^{\circ}$C.

We applied time-correlated single photon counting (TCSPC) to measure exci\-ted-state lifetimes and photoluminescence (PL)
spectra as described elsewhere~\cite{DanielDec04}. 
The excitation source was a pulsed diode laser (PicoQuant LDH400, 20\,MHz, 70\,ps FWHM, 407\,nm (3.05\,eV)). The luminescence was detected with a microchannel plate photomultiplier (Hamamatsu) coupled to a spectrometer and TCSPC electronics (Edinburgh Instruments Lifespec-ps and VTC900 PCI card). 
The temporal resolution is close to 80\,ps, while the spectral resolution is around 4\,nm. The absorption spectra were measured using a UV-Visible spectrophotometer (Varian, Carry 300) with a spectral resolution lower than 1\,nm.

\section{Results}\label{sec:Results}

\begin{figure}
\includegraphics[width=7cm]{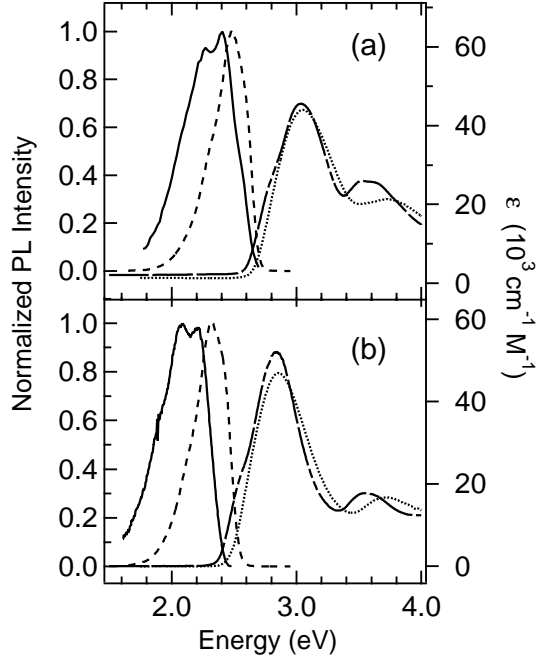}
\caption{\label{F1AbsPL} Absorbance and PL spectra of MOPV3 (a),
and MOPV4 (b). The left axis is the normalised PL intensities in
the dissolved (90$^{\circ}$C, dashed lines) and aggregated phases
(14$^{\circ}$C, continuous lines). The right axis is the decadic
molar extinction coefficient in the dissolved (90$^{\circ}$C,
dotted lines) and aggregated phases (14$^{\circ}$C, long-dash
lines). The photon energy at which time-dependent photoluminescence intensity was measured for the described kinetic analysis is indicated by the arrows.}
\end{figure}

\begin{figure}
\includegraphics[width=7cm]{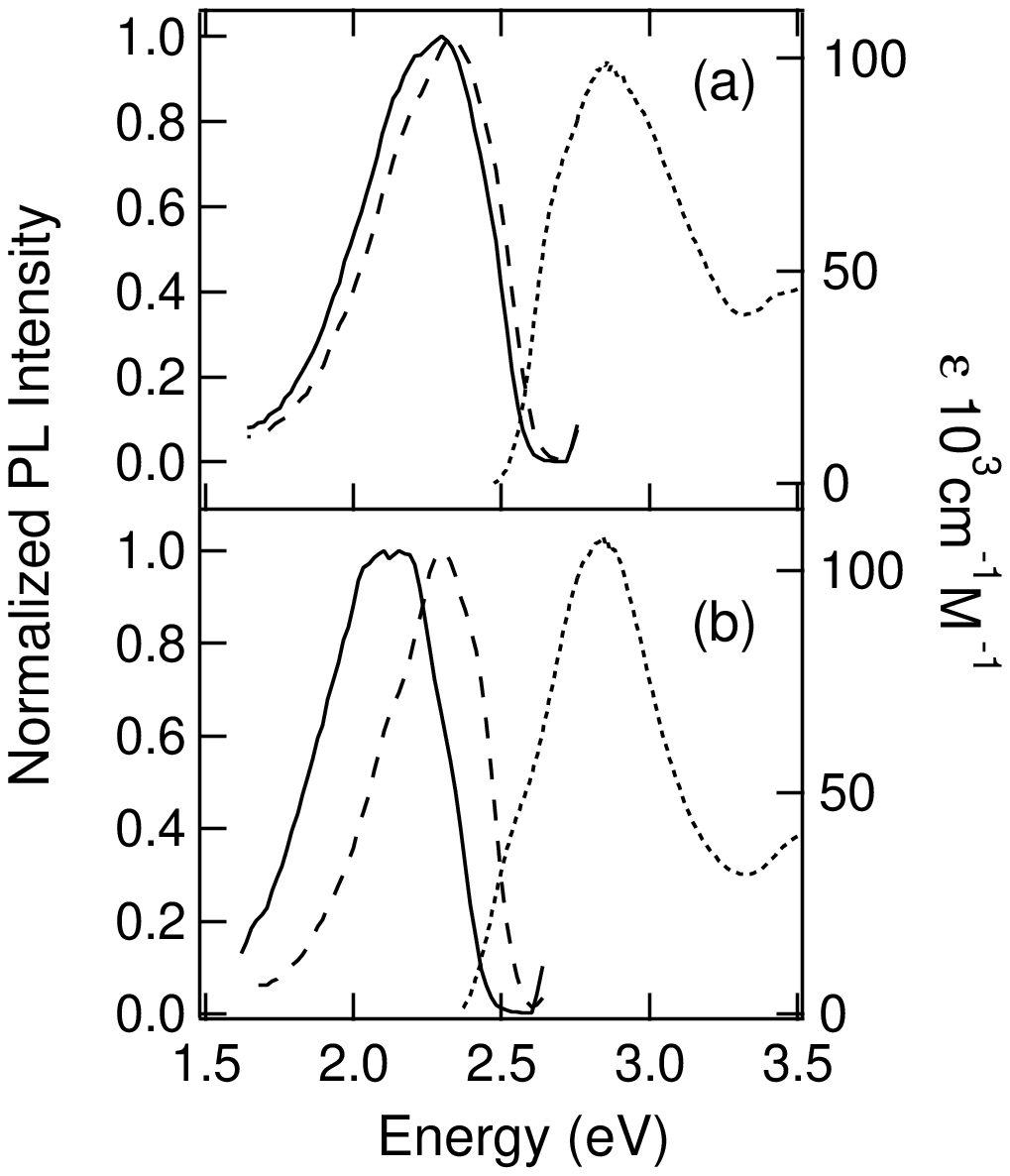}
\caption{\label{BOPVspec} Absorbance and PL spectra of BOPV3 (a)
and BOPV4 (b). The left axis is the normalised PL intensities in
the dissolved (90$^{\circ}$C, dashed lines) and aggregated phases
(14$^{\circ}$C, continuous lines). The right axis is the decadic
molar extinction coefficient in the aggregated phases
(25$^{\circ}$C, dotted lines). The photon energy at which time-dependent photoluminescence intensity was measured for the described kinetic analysis is indicated by the arrows.}
\end{figure}

The absorption and photoluminescence (PL) spectra of MOPV derivatives at 14$^{\circ}$C and 90$^{\circ}$C are shown in Fig.~\ref{F1AbsPL}. 
A red shift of the PL ($\sim 0.2$\,eV) and the appearance of a new absorption shoulder in the red edge of the main band are observed upon cooling the solutions and are attributed to the formation of supramolecular assemblies and to inter-chromophore coupling~\cite{Herz03,Schenning01}. The absorption and PL spectra of BOPV derivatives at 14$^{\circ}$C and 90$^{\circ}$C are shown in Fig.~\ref{BOPVspec}.
We observe a red shift of the PL upon cooling the solutions ($\sim 0.2$\,eV for BOPV4 but smaller for BOPV3) and a red shoulder in the absorption spectra. By analogy with MOPV derivatives, they are attributed to the formation of supramolecular assemblies and to inter-chromophore coupling.

In MOPV and BOPV stacks, RET involving nearest-neighbour interactions are not adequately described with F\"{o}rster theory due to the breakdown of the point-dipole approximation resulting from the non-negligible size and shape of the excited-state wavefunctions compared to the donor-acceptor separation~\cite{Beljonne02,Hennebicq05}. However, at sufficiently low acceptor mole fraction and at low exciton densities, and if \emph{homo}transfer (i.e.\ exciton diffusion) dynamics are negligible, then RET processes can be described with a F\"{o}rster model since on average the donor-acceptor separation is large. With this approximation, a one-step F\"{o}rster model predicts a time dependence of the excitation transfer rate of $t^{(\Delta/6)-1}$, with $\Delta$ being the dimensionality of the acceptor distribution. This result is the generalisation of the methodology developed by Eisenthal and Siegel~\cite{Eisenthal64} for three-dimensional RET for a situation with arbitrary dimensionality. The time-dependent population of the donor exciton density, $n$, after pulsed photoexcitation, is governed by the following rate equation.
\begin{equation}\label{eq:DonorRateEq}
    \frac{d}{dt}n(t) = g \left(t\right) - \frac{n(t)}{\tau} - \gamma t^{\left(\frac{\Delta}{6}-1 \right)} n(t)
\end{equation}
Here $g(t)$ is the exciton generation function, $\tau$ is the excited-state lifetime of the donor in the absence of acceptors,
and $\gamma$ is the rate constant for RET. If the excitation pulse is very short compared to the characteristic timescales of $\tau$ and $\gamma$, we may approximate $g(t) = n_{0} \delta(t)$, where $n_{0}$ is the $t = 0$ exciton density of the donor. The time-dependent donor population density is then given by
\begin{equation}\label{eq:DonorPopulation}
    n(t) = n_{0} \exp \left( -\frac{t}{\tau} - \frac{6\gamma}{\Delta} t^{\Delta/6} \right)
\end{equation}
with $\gamma$ given by
\begin{equation}\label{eq:FoersterRateConst}
\gamma = R^{\Delta} \rho \frac{\Delta\,\pi^{\Delta/2}\,
\Gamma \left(1-\Delta/6\right)}{6\,\Gamma\left(1+\Delta/2\right)\,
\tau^{\Delta/6}}
\end{equation}
where $\Gamma$ is the gamma function, $R$ the F\"{o}rster radius and $\rho$ is the acceptor density in $\Delta$ dimensions with units m$^{-\Delta}$. We thus find that within a generalized F\"orster model, the time-dependent population decay should follow a stretched exponential function where the stretching parameter depends on the dimensionality of the transfer process.

\begin{figure}
\includegraphics[width=7cm]{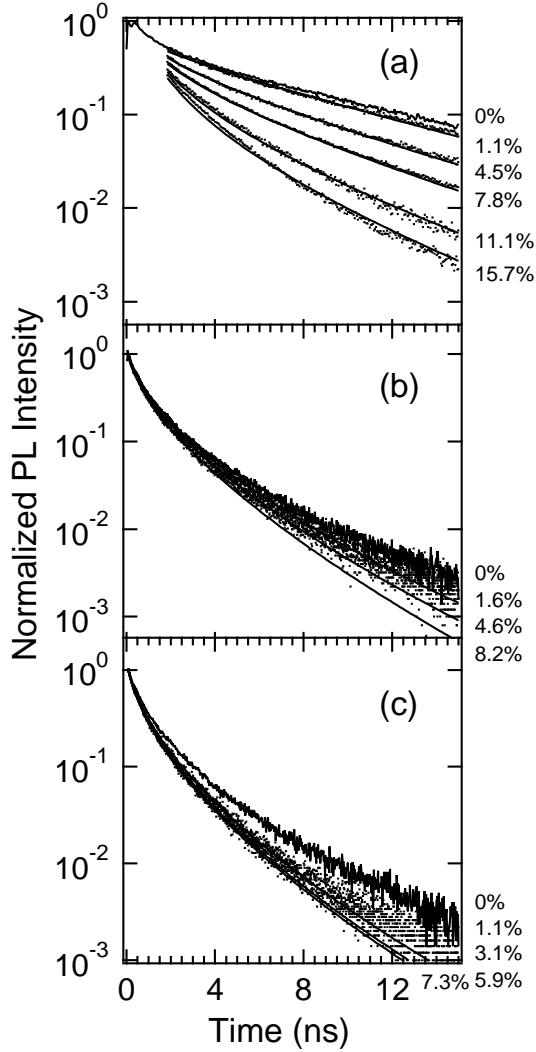}
\caption{\label{fig:TCSPCdecays} PL intensity decay of three
blends measured at a photon energy where only the donor emits and
at 14\,$^{\circ}$C: MOPV3/MOPV4 (2.61\,eV), BOPV3/MOPV4 (2.64\,eV)
and BOPV3/BOPV4 (2.64\,eV). The donor concentrations were kept
around $8 \times 10^{-5}$\,M while the mole fractions of the
acceptors were varied as indicated in the figure. The lines
through the data result from a global fit to $I(t)$
(equation~\ref{eq:StretchExpFit}) in the time window spanning 0 to
20\,ns, see text.}
\end{figure}

To investigate the influence of supramolecular assembly on RET, we have studied three series of blends: MOPV4 in MOPV3, BOPV4 in MOPV3 and BOPV4 in BOPV3, where in each case the short oligomer is the energy donor and the long oligomer is the energy acceptor. As the laser excites both materials, we probe only the decay of the donor (at 2.61 eV for MOPV3 and 2.64\,eV for BOPV3, indicated by the arrows in Figs.~\ref{F1AbsPL} and \ref{BOPVspec}) and measure the enhancement of the decay as the mole fraction of the acceptor increases from 0\% to
$\sim 15$\%. Fig.~\ref{fig:TCSPCdecays} displays the PL decay kinetics  at these detection photon energies of various blend solutions with mole fraction of acceptor ranging from 0\% to 10\%. 

BOPV3 displays non-exponential decay kinetics over all timescales investigated here~\cite{DanielDec04}, while MOPV3 displays exponential decay kinetics after $\sim 2$\,ns (the time window that was used for the fit procedure). In order to extract $R$ and $\Delta$ from the data of the three blends, we first fitted the MOPV3 and BOPV3 decays from the 0\% mole fraction solutions with a stretched-exponential function ($I(t) = a \exp (- kt^{-h})$). The results were $k=0.126$\,ns$^{-1}$ and $h=1$ for MOPV3 and $k=1.38$\,ns$^{-0.6}$ and $h=0.6$ for BOPV3. Note that $h=0.6$ can be related to diffusion-assisted exciton-quenching at defects on a three-dimensional lattice~\cite{Herz03,DanielDec04}.

We then applied a global fit to PL decays at all different MOPV4 and BOPV4 mole fraction with
\begin{equation}\label{eq:StretchExpFit}
    I(t) = a \exp \left( - kt^{-h} - bt^{-c} \right)
\end{equation}
where $k$ and $h$ were fixed to the values found in the undoped nanostructures, $a$ and $b$ were allowed to float for each individual data set, and $c$ was only allowed to float globally for the entire data set (see Fig.~\ref{fig:TCSPCdecays} for the results).

For the MOPV3/MOPV4 blends, the best global fits yield $c = 0.21 \pm 0.01$ which corresponds to a dimension $\Delta = 1.3 \pm 0.1$. If we constrain the value of $c$ to 0.5 (for a three-dimensional acceptor distribution), the goodness-of-fit deduced by statistical analysis of the $\chi^{2}$ values is at least a factor of two worse than if $c$ is allowed to float freely. For the BOPV3/BOPV4 blends, the situation is reversed and the best global fits yield $c=0.52 \pm 0.01$ which correspond to a dimension $\Delta = 3.1 \pm 0.1$. Constraining the value of $c$ to 0.17 (for a one-dimensional acceptor distribution), reduced the goodness-of-fit by a factor two. For the BOPV3/MOPV4 blends, the situation is less clear as the global fits converge to a non-physical value of $c \approx 1$. If the value of $c$ is constrained to $c = 0.5$ for a three-dimensional distribution, the goodness-of-fit does not decrease significantly ($\chi^2$ does not change), while if the value of $c$ is constrained to one or two-dimensional acceptor distribution, the fit quality becomes poor ($\chi^2$ increases by more than a factor of three).

\begin{figure}
\includegraphics[width=7cm]{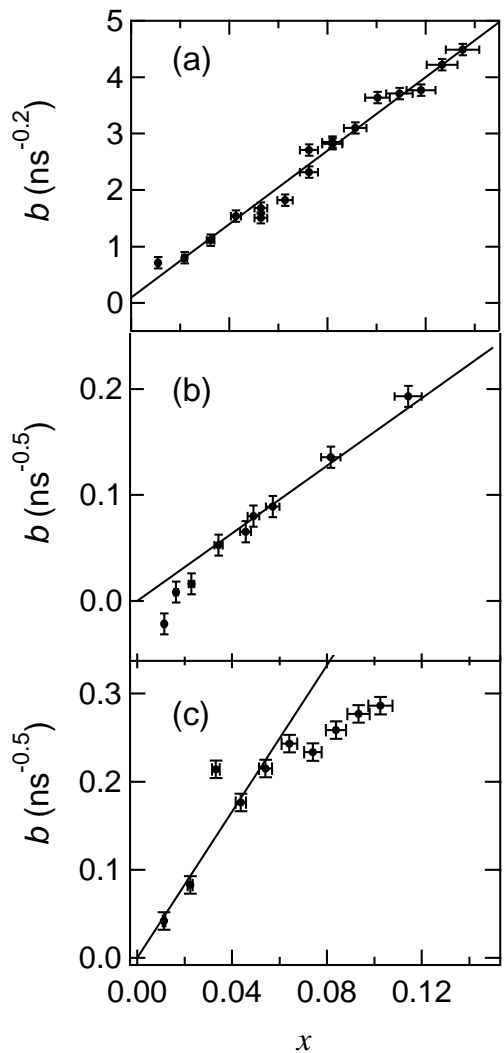}
\caption{\label{fig:FitParamsPlot} Fitting coefficients $b$ versus
the acceptor mole fraction $x$ for the MOPV3/MOPV4 (a),
BOPV3/MOPV4 (b) and BOPV3/BOPV4 (c) blends.}
\end{figure}

To extract the F\"{o}rster radius $R$ in the three configurations, we plot $b$ versus $x$, where $x$ is the acceptor mole fraction. In a one-dimensional distribution, the acceptor concentration in the stack, $\rho_{\Delta=1}$, is defined as $x / \overline{r}$ with $\overline{r} = 0.35$\,nm the average intermolecular separation~\cite{Hoeben03}. In a three-dimensional distribution, the acceptor concentration, $\rho_{\Delta=3}$, is defined as $x / \overline{v}$ with $\overline{v}$ the average molecular volume. We approximate this volume with a cylinder section of height 0.7\,nm (as BOPV derivatives consist of two oligomers) and radius 2.5\,nm (the experimental radius of the supramolecular stacks). From the slopes of the plots $b$ versus $x$ shown in Fig.~\ref{fig:FitParamsPlot}, we obtain the F\"{o}rster radii of $R$ = 7.8, 2.3 and 1.6 \,nm respectively for the blends MOPV3/MOPV4 (one-dimensional distribution), BOPV3/BOPV4 (three-dimensional distribution) and BOPV3/MOPV4 (three-dimensional distribution). Note that as the coefficients $b$ for the BOPV3/BOPV4 blends saturate above 7\% mole fraction (probably due to phase segregation effects), only the first part of the curve was used to determine the F\"{o}rster radius. 
Given that the interchromophore co-facial distance is 3.5\,\AA~\cite{Beljonne05}, this indicates that RET from MOPV3 to MOPV4 is competitive with all other de-excitation processes over a distance spanning up to 22 oligomers \emph{primarily along the stack direction}, whereas in the BOPV structure this process competes for donor-acceptor separations equivalent to approximately 7 oligomers away, both long the stack \emph{and} across to the opposite helix. Because the stacks are typically hundreds of nanometers in length, the distance scales extracted here are reasonable.

\section{Discussion}

The model used to extract F\"{o}rster radii from the PL decay of the blends assumes that multi-step homotransfer dynamics in the donor architecture are negligible and that the acceptor mole fraction is sufficiently low so that, on average, the donor-acceptor separation is longer than the nearest-neighbour separation to avoid the complications imposed by the break-down of the point-dipole approximation~\cite{Beljonne02,Hennebicq05}. We consider the functional form of the PL decay rate of the undoped nanostructures to explore these conditions. In a previous publication describing femtosecond-resolved transient PL measurements in MOPV4~\cite{Herz03}, we found that a stretched exponential function of the form $I(t) = I_{0} \exp \left( -t/\tau \right)^{\beta}$ describes the PL decay of the supramolecular assemblies. In the stacked phase, $\beta = 1/3$ over the first 600\,ps. We invoked models relating $\beta$ to the lattice dimensionality $d$ by $\beta = d/(d + 2)$. (Note that $d$ and $\Delta$ discussed here have slightly different meaning; the dimensionality of the lattice in which excitons undergo multiple incoherent hops during their lifetime is $d$, whereas here $\Delta$ is the dimensionality of the donor-acceptor distribution in one-step transfer processes.)
We thus argued that multistep exciton diffusion in a quasi-one-dimensional lattice is a plausible description of exciton dynamics in MOPV over this short timescale. At longer times ($>2$\,ns) we invoked a higher dimensionality of the donor-acceptor distribution, as excitons located in a local minimum of the potential energy landscape need to interact with suitable transfer sites that are located further away and the probability of transfer to sites in the opposite helix of the architecture is non-negligible. This is a picture that is also consistent with MOPV3 stacks; the pure MOPV3 data in Fig.~\ref{fig:TCSPCdecays} display non-exponential decay at early time, switching to exponential decay after a few nanoseconds as reported in the case of MOPV4~\cite{Herz03}, when the excitation is no longer mobile and simple radiative and non-radiative pathways of the localised excitons dominate the decay.

The situation changes upon addition of deeper traps in the form of MOPV4 to MOPV3 stacks, where PL decay on nanosecond timescales becomes non-exponential again due to RET and the distribution of suitable acceptor sites displays quasi-one-dimensional characteristics once again. Localised excitons in MOPV3 undergo \emph{single} step transfer and see a predominantly one-dimensional distribution of MOPV4. This process is efficient indicated by the large value of $R$ ($\sim 8$\,nm). The donor-acceptor spectral overlap is similar in MOPV and BOPV stacks, so the increased efficiency in the MOPV system cannot be explained with more favourable resonance conditions. We rationalise the high efficiency in MOPV stacks as due to increased order and periodicity in the MOPV architectures. The likelihood of finding acceptors with favourable orientations is high along the stack. Adjacent co-facial oligomers are displaced by an angle of 12$^{\circ}$ and the chromophore consists of two oligomers on average due to the moderate intermolecular coupling energies~\cite{Spano07}. Therefore the next chromophore with the same orientation to any given photoexcited chromophore is roughly 8 chromophores away on average. Over a distance covering 22 chromophores, corresponding to $R$, any photoexcited chromophore would see roughly 3 chromophores with the same orientation, so around that exciton an acceptor occupying at least those 6 sites in total would face a high probability of RET (and in practice many more sites have an important projection along the same orientation as the donor). If this periodicity is not present, however, as in the case of BOPV, then the probability of finding an acceptor with a significant projection along the axis of the transition dipole moment of the donor is more limited along the stack and is comparable to that across to the other helix, rendering the donor-acceptor probability distribution more three-dimensional.

We have established that RET from MOPV3 hosts to MOPV4 guests in mixed supramolecular stacks of the two oligomers is efficient. The picture emerging from Section~\ref{sec:Results} is the following. At low MOPV4 mole fraction, a significant extent of RET occurs within the first $\sim 100$\,ps after absorption of light by MOPV3. This is consistent with our previous report of ultrafast PL depolarization in these mixed nanostructures~\cite{Chang06}. During this time, significant exciton diffusion occurs in MOPV3~\cite{Herz03}, which assists exciton transfer to MOPV4. Localised excitons in MOPV3 undergo at later times \emph{single} step transfer and see a predominantly one-dimensional distribution of MOPV4. Excitation diffusion is still significant in the BOPV3 host nanostructures over these nanosecond windows~\cite{Chang06}, but, on average, the donor-acceptor separation is still large in order to satisfy the conditions of the F\"{o}rster model, especially at low concentrations ($< 5$\%) where $b$ is found to be linear.  The global fitting procedure applied in this paper strongly points to a dependence of the dimensionality of the RET process on the morphology of the supramolecular nanostructures.  For \emph{localised} excitons~\cite{Spano07} characteristic of the long timescales investigated here, the most ordered host structure (MOPV3) displays one-dimensional energy transfer, whereas both blends with the more disordered host structure (BOPV3) display three-dimensional energy transfer. These results point to the importance of controlling supramolecular structure in optimising electronic processes in these types of nanostructures. In this context, the optimisation consists of enhancing the long-range RET efficiency along a specific direction by inducing supramolecular order. This would produce a means to funnel energy uni-directionally to desired exciton dissociation centres over long timescales in photovoltaic applications, for example.

This scenario for MOPV3 host structures appears to be distinct from that invoked to describe exciton bimolecular annihilation processes in MOPV4~\cite{Daniel03}. In that case, a bimolecular annihilation rate constant with explicit time dependence in the form $t^{-1/2}$ was required to reproduce femtosecond transient absorption data at high pump fluences ($\geqslant 100$\,$\mu$J\,cm$^{-2}$). We interpreted this as indicative of a non-Markovian exciton bimolecular depletion mechanism mediated by long-range RET interactions. In contrast to the analysis presented here, an effective three-dimensional exciton distribution was deduced from the exponent of the time dependence of the bimolecular annihilation rate constant. We reconcile this with the analysis presented here by pointing out that the bimolecular annihilation process occurs in a picosecond timescale where the exciton diffusivity is high and sites close to acceptor sites in the opposite helix can be reached more readily by multiple hops, rendering the apparent acceptor distribution to have a higher dimensionality than one. We pointed out that for this reason, a microscopic description is more adequate to describe the bimolecular annihilation phenomena~\cite{Beljonne05,Daniel07}.

A quantitative description of the processes in this fast (sub-nanosecond) time\-scale is beyond the scope of this paper. Firstly, it would require a statistical treatment of microscopic events within the mixed MOPV stack~\cite{Beljonne05,Daniel07}. Secondly, an appropriate description of the donor-acceptor electronic coupling is more complex~\cite{Spano07}. 
A full representation of the RET dynamics in this situation requires a model that goes beyond F\"{o}rster theory in the description of the intermolecular electronic coupling~\cite{Beljonne02,Hennebicq05} and, depending on the magnitude of this coupling, perhaps away from the golden rule rate expression derived from second-order perturbation theory~\cite{May00}.

\section{Conclusion}
We have explored the photophysical consequences of supramolecular assembly of oligo-\emph{p}-phenylenevinylene derivatives in dilute solution. We have shown that the supramolecular assemblies favour the funnelling of the energy through resonance energy transfer (RET). RET can be modelled on the nanosecond timescale with a F\"{o}rster formalism but the effective rates depend strongly on the exact stack configuration. As BOPV derivatives are less ordered than MOPV derivatives, the RET rates are smaller and the dimensionality of this process increases from one. However, BOPV assemblies offer a promising route to solid state supramolecular assembly~\cite{Hoeben05}.

Our results show clearly that control of order in the nanometre lengthscale provides a promising strategy for harvesting energy in supramolecular semiconductor systems. MOPV and BOPV derivatives represent a very good model system to study these effects as they possess polymeric optoelectronic properties in the aggregated phase but with the additional tunability and structural control afforded by supramolecular chemistry.

\section{Acknowledgements}
CS acknowledges support from NSERC and the Canada Research Chairs Programme. The Work in Eindhoven is supported by the Netherlands Organisation for Scientific Research (NWO, CW). The Cambridge-Eindhoven collaboration was supported by the European Commission (LAMINATE).



\end{document}